%% file: main.tex
\documentclass[a4paper,fleqn]{cas-sc}

\usepackage[numbers,sort&compress]{natbib}
\usepackage{slashed}

\def\tsc#1{\csdef{#1}{\textsc{\lowercase{#1}}\xspace}}
\tsc{WGM}
\tsc{QE}
\tsc{EP}
\tsc{PMS}
\tsc{BEC}
\tsc{DE}
\begin{document}

\title[mode = title]{Fermion-Vortex Interactions in Axion Electrodynamics}              
\shorttitle{Fermion-Vortex Interactions in Axion Electrodynamics}
\shortauthors{S. Kantha et~al.}

\author[1,2]{Saurav Kantha}[orcid=0009-0008-5559-5044]
\ead[skantha@irsamc.ups-tlse.fr]{skantha@irsamc.ups-tlse.fr}

\affiliation[1]{organization={Department of High Energy and Astrophysics, 
S.N. Bose National Centre for Basic Sciences},
                addressline={JD Block, Sector III}, 
                city={Kolkata},
                state={WB},
                country={India}}

\author[1]{Amitabha Lahiri}[
           orcid=0000-0001-8113-6345,   % <-- ORCID goes here
           % you may add more keys such as:
           % type=editor,  role=Researcher,  auid=000,  bioid=1
]
%[type=editor]
\ead[amitabha@bose.res.in]{amitabha@bose.res.in}
\cormark[1]

\affiliation[2]{organization={Laboratoire de Physique Théorique,
Université Paul Sabatier},
                addressline={118 Route de Narbonne}, 
                postcode={31400}, 
                city={Toulouse},
                country={France}}

\cortext[cor1]{Corresponding author}

\begin{abstract}
A relativistic action for scalar condensate-fermion mixture is considered where both the scalar boson and the fermion fields are coupled to a $U(1)$ gauge field. The dynamics of the gauge field is governed by a linear combination of the Maxwell term, and the Lorentz invariant $\mathbf{E\cdot B}$ term with a constant coefficient $\theta$.
We obtain an effective action describing an emergent fermion-fermion interaction  and fermion-vortex tube interaction by using the particle-string duality, and find that the $\theta$ term can significantly affect the interaction of fermions and vortices. We also  perform a dimensional reduction  to show a $\theta$ dependent flux attachment to the itinerant fermions.
\end{abstract}

% \begin{graphicalabstract}
% \includegraphics{figs/cas-grabs.pdf}
% \end{graphicalabstract}

% \begin{highlights}
% \item Research highlights item 1
% \item Research highlights item 2
% \item Research highlights item 3
% \end{highlights}


\maketitle

\input{introduction}

\input{Model}

\input{dualization}

\input{fermion_interactions}

\input{dimensional_reduction}

\input{summary}

\input{acknowledgements}

\bibliographystyle{unsrt}

\bibliography{ref.bib}

\end{document}

%% file: introduction.tex
\section{Introduction}

The interaction of charged currents with rank-1 gauge fields is a well understood concept in physics, both in high-energy and condensed matter sectors. In relativistic physics, the coupling goes by the name of minimal coupling, whereas in condensed matter physics, the gauge fields are introduced on the lattice via Peierls substitution \cite{peierls1933zur}. In both cases, the gauge invariance is ensured by a local phase change of the  states as the gauge field $A$ goes to $A + d\phi$ under the action of the underlying gauge group. 

However, as we move one rank higher in the gauge field i.e. from a rank-1 gauge field $ A_\mu$ to a rank-2 gauge field $ B_{\mu\nu}$, a gauge-invariant local coupling of fermions to the gauge field  is no longer possible. A gauge-invariant, but nonlocal, interaction between fermions and an antisymmetric rank-2 tensor gauge field was recently found in the context of topological mass generation~\cite{choudhury2015topological} . In this interaction, the gauge field $B_{\mu\nu}$ couples to the fermion current $j^\mu$  via the nonlocal term 
$B_{\mu\nu}\tilde{J}^{\mu\nu}$, where $\tilde{J}^{\mu\nu} = \epsilon^{\mu\nu\rho\lambda}\frac{1}{\Box}\partial_\rho j_\lambda$ is a nonlocal current because of the appearance of the inverse  of the four-dimensional wave operator $\Box\equiv \partial_\mu\partial^\mu$.
For the purposes of calculations, we can think of $\frac{1}{\Box}$ as a Green function, or in momentum space as $-\frac{1}{k^2}\,.$ 

The consequences of such a coupling have not been explored very much.  It has been known for a long time that the rank-2 antisymmetric gauge field $B_{\mu\nu}$ mediates interactions between strings~\cite{Kalb:1974yc}. They have in the recent past also found applications in the context of various field theoretic dualities~\cite{Murugan:2021jwu} and in condensed matter physics, in the description of bosonic topological insulators~\cite{moore2011, Vishwanath2013, obliueTopoIns, tiwari2014topological}.

In this work, the presence of charged fermions gives rise to the nonlocal interaction.  The non-relativistic limit of the nonlocal current $\tilde{J}^{\mu\nu}$ includes the spin density current of the fermions, so $B_{\mu\nu}$ can be thought of as a gauge field for spin. 
It was shown recently that such couplings could result in interesting physics such as a linear attractive potential between fermions~\cite{mukherjee2022spin},  and attachment of magnetic flux to fermions resulting in anyonic statistics~\cite{mukherjee2023spin}. The flux-attached fermions in such a system are essentially bound states of topological vortices and electrons. Bound states between topological textures and electrons, specifically skyrmion-electron bound states have been recently studied in~\cite{revaz_2,revaz,pujol}. The field theoretical formalism used in this paper to study vortex-electron interactions can be adapted to the study of such systems.  

 There is another aspect of gauge fields, which has received a lot of attention in the past two decades, particularly in the context of Topological Insulators (TIs) and Topological Superconductors (TSCs). This aspect is the topological action of the gauge field~\cite{QiEffectiveTheory2008,theta1doi:10.1126/science.aaf5541,theta2sekine2021axion,theta3doi:10.1073/pnas.1515664112,theta4PhysRevB.89.054506,theta5Martín-Ruiz2021,theta6PhysRevB.103.235111,theta7PhysRevB.93.045115,theta8PhysRevB.87.085132,axion_mie_theory,ryu_anomaly}. In 3+1 dimensions, there are two possible dynamical terms of a rank-1 gauge field. One of them is the usual Maxwell term $-\frac{1}{4}F^{\mu\nu}F_{\mu\nu}$ and the other one is the so called axion term or the $\theta$-term, usually written as $\frac{\theta}{32\pi^2} \epsilon_{\mu\nu\lambda\rho} F^{\mu\nu}F^{\lambda\rho}$ which is topological, meaning that it does not affect the local dynamics of the gauge field . In particular, the importance of the axion term has been highlighted in~\cite{QiEffectiveTheory2008} where the authors have established that the axion term is responsible for the quantized electromagnetic response of time-reversal invariant topological insulators in 3 spatial dimensions. 
%  This quantized electromagnetic response is described in terms of the surface conductivity of a three dimensional block of a topological insulator 
% %
% \begin{equation}
%  \sigma_{xy} =    \int_{z_1}^{z_2}\frac{d\theta}{2\pi}
% \end{equation}\label{hall_theta}
% %
% showing that there is a Hall effect is induced in the intersection of two regions with different values of $\theta$. 
A recent work~\cite{Murugan:2021jwu} also explores the dualities between these two the field theories describing 3+1 dimensional TIs and TSCs and their parallels to the particle-vortex duality~\cite{Murugan2017}. The axion term has also been found to be important systems like  semimetals \cite{Weyl1,Weyl2Gooth2019,Weyl3SCHMELTZER2020168237,Weyl4mottola2023axions,Weyl5PhysRevLett.111.027201,Weyl6PhysRevB.87.161107}.

% {\color{black}
% In one lower dimension, a topological field theory which gives rise to many interesting phenomena is the Chern-Simons theory,
% %
% \begin{equation}
%     \frac{e^2C_1}{f\pi \hbar}\int d^3x \epsilon^{\mu\nu\lambda}A_\mu\partial_\nu A_\lambda\,.
% \end{equation}\label{chern_simons}
% %
% A particle current coupled to a Chern-Simons theory produces states of fractional statistics through the attachment of fluxes to particles~\cite{Wilczek:1981du}. The Laughlin wavefunction~\cite{Laughlin:1983fy}, which provides an excellent description of the ground state of systems showing fractional quantum Hall effect (FQHE), supports quasiparticles with fractional charge and fractional statistics, which can also be thought of particles attached to fluxes. This similarity has led to the suggestion that FQHE may have an effective description in terms of an emergent Chern-Simons gauge theory~\cite{Fradkin:1991wy}. 
% }

In this paper, we will be interested in systems which include features of both of the aforementioned elements, i.e. a rank-2 gauge field $B_{\mu\nu}$ and the axion term for a rank-1 gauge field. One example of such a system is a construction involving two interfaces, proposed in a recent work by Nogueira et al.~\cite{nogueiraWittenEffect, nogueiraDuality} for the detection of the Witten effect~\cite{WITTEN1979283}. In their work, the authors consider a superconductor-topological insulator (TI)-superconductor junction with vortex flux tubes within the two type II superconductors end on the TI surface where there is a change in the value of axion parameter $\theta$, leading to a Josephson-Witten effect. 

To this end, we will begin with a theory of a mixture of bosons and fermions in the background of a linear combination of the Maxwell and axion terms for a rank-1 gauge field, where the  coupling parameter $\theta$ of the axion term is taken to be a constant. When the axion parameter $\theta$ is space-time independent, the axion term in the action is a boundary term and therefore one expects it to not play a role in local interactions. In this work, we show that in the presence of flux-tubes in the background, $\theta$ does affect local interactions.

Theories involving mixtures of bosons and fermions   have been proposed in various contexts including high-$T_c$ superconductivity \cite{BF_1SALAS201637,BF_2PhysRevB.81.064514,BF_3PhysRevB.66.104516,BF_4RANNINGER1995279} to BCS-BEC crossover\cite{BF_BCS_BEC1PhysRevD.76.034013,BF_BCS_BEC_2PhysRevB.102.144506}. In terms of the appearance of flux tubes, our construction bears some resemblance to the Friedberg-Lee model~\cite{friedberg1989gap, Friedberg:1990eg}, proposed as a description of high-$T_c$ superconductivity. We consider the Abelian Higgs model with the presence additional itinerant fermions minimally coupled to the $U(1)$ gauge field. In this theory, vortex flux tubes appear as a consequence of a $U(1)$ symmetry breaking of the bosonic field. In the symmetry broken phase, the theory can be dualized, using a 3+1 dimensional adaptation of the particle vortex duality, into another theory comprised of flux tubes and a rank-2 gauge field mediating the flux tube-flux tube interaction.
The effective interactions between fermions and flux tubes can then be studied by integrating out first the rank-1 gauge field and then the rank-2 gauge field from the partition function describing the system. We do this systematically by bringing the Lagrangian to linear order in the rank-1 gauge field $A_\mu$. This is achieved by multiplying the partition function of the system by a Gaussian path integral comprised of the Maxwell term $F_{\mu\nu}$, its dual and a new rank-2 antisymmetric field $\chi_{\mu\nu}$. After linearizing the Lagrangian, the gauge field $A_\mu$ is integrated out and the antisymmetric field $\chi_{\mu\nu}$ is traded away for various nonlocal gauge invariant couplings between the gauge field $B_{\mu\nu}$ and the fermion current. The gauge field is then integrated out to get the effective interactions. Finally, we perform a dimensional reduction by compactifying the length of the system in the $z$ direction to obtain a planar action and the resultant interactions in an effective 2+1 dimensional system. 

The paper is organized as follows. In section \ref{sec:model}, we start with the
partition function of the system described in terms of a rank-1 gauge field, a scalar condensate and itinerant fermions and and show its dual transformation to another partition function in terms of flux tubes and the Kalb-Ramond gauge field~\cite{Kalb:1974yc} along with the usual rank-1 gauge field.
$A_{\mu}$. In section~\ref{sec:dual}, we linearize the Lagrangian in the rank-1 gauge field to obtain a partition function in terms of the rank-1 gauge field and integrate it out to obtain a partition function with the rank-2 gauge fields, fermions and flux tubes.
In section \ref{sec:eff.int}, the gauge field $B_{\mu\nu}$ is integrated out to obtain
the relevant interactions where it is shown that the axion parameter $\theta$ appears in all local interactions.
In the subsequent section we perform a dimensional reduction to an effective 2+1 dimensional system and demonstrate flux attachment to fermions and finally conclude with a summary and outlook.

%% file: Model.tex
\section{The Model}
\label{sec:model}
%%%%%%%%%%%%%%%%%%%%%%
We consider a modified version of the Abelian Higgs model { , which includes the } parity violating $\theta$ term  { or} the axion term $\theta\epsilon^{\mu\nu\alpha\beta}F_{\mu\nu}{F}_{\alpha\beta}$ alongside the usual Maxwell term $F_{\mu\nu}F^{\mu\nu}$.  Additionally, we also consider the presence of Dirac fermions in the system which are also minimally coupled to the gauge field $A_\mu$\,. { We work with 3+1 dimensional relativistic quantum field theory, even though the
objects and systems we consider are all non-relativistic. There are two important reasons for this approach. One is that the procedure of dualization, which changes the effective variables from the scalar field to the flux tubes and the anti-symmetric tensor gauge field, works best in four dimensions. The other is that electron spin appears naturally in the theory and we do not need to put in an interaction by hand. The coupling of electron spin to the anti-symmetric gauge field then appears automatically from the dualization.}

To this end, we start with the  Lagrangian
%
% The simultaneous consideration of the Maxwell term and theta term as a linear combination is a well-established concept in the literature. Ideally, the dynamics of a one-form gauge field in a system can be expressed as a linear combination of these terms. However, the topological nature of the $\theta$ term means that it does not contribute to the dynamics of the gauge field in the bulk.  There are several examples of systems where the $\theta$  term plays an important role, none more well known than a topological insulator. In analogy to the case of the description of the integer quantum Hall system by the Chern-Simons theory in $(2+1)$-dimensions, the electrodynamic axion term describes the effective theories of certain classes of topological superconductors. This term breaks gauge invariance in the presence of a boundary and hence leads to a gauge anomaly. Cancelling this anomaly involves introducing a theory on the boundary that restores gauge invariance in the bulk+boundary system. In the context of topological insulators, this boundary theory describes gapless Dirac fermions, leading to a conductive surface. It has been demonstrated that the coefficient $\theta$ can assume only specific discrete values in such systems, characterizing different topological phases based on the system's symmetries.  In a generic system, there are no restrictions on the values of $\theta$ , and as stated above, the dynamics of the gauge field should be described by a linear combination of the Maxwell and the axion term.
%
\begin{align}
    \mathcal{L}=-\frac{1}{4}F_{\mu\nu}F^{\mu\nu}&-\frac{\theta}{32\pi^{2}}\epsilon^{\mu\nu\alpha\beta}F_{\mu\nu}F_{\alpha\beta} +\frac{1}{2}(D_{\mu}\phi)^{\dagger}(D^{\mu}\phi)-V(|\phi|^{2})\notag\\ 
&\qquad\qquad +\bar{\psi}(i\gamma^{\mu}\partial_{\mu}-m)\psi-eA_{\mu}\bar{\psi}\gamma^{\mu}\psi\,,
\label{lagrangian_original_bf}
\end{align}
where $D_\mu=\partial_\mu + iqA_\mu$\,.
%The partition function of the system is given by 
%
%\begin{equation}
%Z=\int\big(\mathcal{D}A_{\mu}\mathcal{D}\phi\mathcal{D}\phi^{\dagger}\mathcal{D}\bar{\psi}\mathcal{D}\psi\big)\exp\Big(i\int d^{4}x\mathcal{L}\Big)\,.
%\label{main_part_f}
%\end{equation}

{ We focus on the broken symmetry phase in which $D_\mu\phi=0$ and there are flux strings carrying quantized magnetic flux}~\cite{nielsen1973vortex}
\begin{equation}
\oint_{C}dx^{\mu}A_{\mu}=-\frac{2n\pi}{q}\,,
\label{flux_quantization}
\end{equation}
where $n\in\mathbb{Z}$ is the winding number. The description of $\phi$ in this
phase is of the form 
\begin{equation}
\phi=vf(|{\bf r}-{\bf r}_{0}|)\exp(i\chi({\bf r}-{\bf r}_{0}))\,,
\label{dual.stringphi}
\end{equation}
where $v$ is the vacuum expectation value of the scalar  condensate $\phi$, while  $f$ is a function of the distance of a point {$\bf r$} from the core of the flux tube ${\bf r}_0$ {with the property that} as $\mathbf{r}\rightarrow \mathbf{r}_0$, $f\rightarrow 0$ and as ${\bf r}\rightarrow\infty$, we have  $f\rightarrow 1\,.$ For $\phi$ to admit a topological solution, its phase $\chi$ must
be a multivalued function. Hence we can split $\chi$ into a well behaved
single-valued regular part $\chi_{r}$ and a singular part $\chi_{s}$
as $\chi=\chi_{r}+\chi_{s}$. The regular part of the phase $\chi_{r}$ can be integrated out by
the introduction of an auxiliary field~\cite{mathur_dual,km_lee_strings,akhmedov_dual}, 
to arrive at the dual Lagrangian
away from the core
\begin{align}
 %   \begin{split}
        Z =& \int\big(\mathcal{D}A_{\mu}\mathcal{D}B_{\mu\nu}\mathcal{D}\bar{\psi}\mathcal{D}\psi\mathcal{D}\chi_{s}\big) \notag\\
        & \exp\Bigg[i\int d^{4}x\Bigg(-\frac{1}{4}F_{\mu\nu}F^{\mu\nu}-\frac{\theta}{32\pi^{2}}
        \epsilon_{\mu\nu\alpha\beta}F^{\mu\nu}F^{\alpha\beta} \notag \\
&\qquad +\frac{1}{12v^{2}}H^{\mu\nu\lambda}H_{\mu\nu\lambda}-\frac{1}{2}\epsilon^{\mu\nu\alpha\beta}B_{\alpha\beta}\partial_{\mu}\partial_{\nu}\chi_{s} \notag \\
%% &{\color{red}-\frac{q}{2}\epsilon^{\mu\nu\alpha\beta}\partial_{\nu}B_{\alpha\beta}A_{\mu}} \notag\\
&\qquad {-\frac{q}{2}\epsilon^{\mu\nu\alpha\beta}H_{\nu\alpha\beta}A_{\mu}}\ +\bar{\psi}(i\gamma^{\mu}\partial_{\mu}-m)\psi-eA_{\mu}\bar{\psi}\gamma^{\mu}\psi\Bigg)\Bigg]\,,
\label{linear_z_A}
%    \end{split}
\end{align}
{where we have written $H_{\nu\alpha\beta} = \frac{1}{3}(\partial_{\nu}B_{\alpha\beta} + \partial_{\alpha}B_{\beta\nu} + \partial_\beta B_{\nu\alpha})$\,.}

Since the scalar field $\phi$ vanishes at the core of the flux tube, { derivatives on the} singular part of its phase $\chi_s$ {are non-commuting. The commutator $\partial_{[\mu}\partial_{\nu]}\chi_s$} is the vorticity  of the flux tube,
{ and the  singularities of this vorticity field lie on the string core. 
It is easy to check that the dual of this field represents the worldsheet swept out by the vortex string core, 
%The ANO string world sheet $\Sigma$ is the collection of singular points of $\theta\,,$ 
%
\begin{equation}\label{string-ws}
\Sigma^{\mu\nu} \equiv \epsilon^{\mu\nu\lambda\rho}\partial_\lambda\partial_\rho\chi_s =  
2\pi n\int d^2\sigma\, \epsilon^{ab}\frac{\partial X^\mu}{\partial \sigma^a} 
\frac{\partial X^\nu}{\partial \sigma^b} \delta^4({x} - {X}(\sigma))\,.
\end{equation}
where we have included
the vorticity quantum $2\pi$ and the winding number $n$ in the definition of the world sheet. }
The integral of the vorticity along a closed loop $C$ vanishes if $C$ does not contain the core of the vortex string.

% \newpage

%% file: dualization.tex
\section{Dualization and Emergent Couplings}
\label{sec:dual}
%%%%%%%%%%%%%%%%%%%%%%%%%%%%%%%%%%
We start by linearizing the gauge field $A_\mu$ in the Lagrangian { through the introduction of}
an anti-symmetric tensor $\chi_{\mu\nu}$ . To this end, we { multiply the partition function of Eq.~(\ref{linear_z_A})} 
by the following Gaussian functional 
\begin{equation}
Z_{\chi}=\int D\chi_{\mu\nu}\exp\Big(-i\int d^{4}x\big(\alpha F^{\mu\nu}+\frac{\beta}{2}\epsilon_{\mu\nu\alpha\beta}F^{\alpha\beta}+\frac{1}{2}\chi_{\mu\nu}\big)^{2}\Big)\,,
\label{dual.Z_chi_mu_nu_gaussian}
\end{equation}
{ which has a constant value,} and set 
\begin{align}
\alpha^{2}-\beta^{2}  =-\frac{1}{4}\,, &\\
\alpha\beta=-\frac{\theta}{32\pi^{2}}\,.
\label{dual.alpha_beta_soln}
\end{align}
{ For this choice of $\alpha$  and $\beta$\,, the terms quadratic in $F_{\mu\nu}$ 
cancel out in the product. We are then left with the following expression for the partition function:}
% {The product is the partition function}
%
\begin{align}
   % \begin{split}
        Z_{dual}&=\int\big(\mathcal{D}\chi_{\mu\nu}\mathcal{D}A_{\mu}\mathcal{D}B_{\mu\nu}\mathcal{D}\bar{\psi}
        \mathcal{D}\psi\mathcal{D}\chi_{s}\big)\notag \\
        &\qquad \exp\Bigg[i\int d^{4}x\Bigg(-\frac{1}{4}\chi_{\mu\nu}\chi^{\mu\nu}-\alpha F^{\mu\nu}\chi_{\mu\nu}-\frac{\beta}{2}\epsilon_{\mu\nu\alpha\beta}\chi^{\alpha\beta}F^{\mu\nu}\notag \\
&\qquad\qquad\qquad  +\frac{1}{12v^{2}}H^{\mu\nu\lambda}H_{\mu\nu\lambda}-\frac{1}{2}\epsilon^{\mu\nu\alpha\beta}B_{\alpha\beta}\partial_{\mu}\partial_{\nu}\chi_{s}\notag \\
&\qquad\qquad\qquad  -\frac{q}{2}\epsilon^{\mu\nu\alpha\beta}\partial_{\nu}B_{\alpha\beta}A_{\mu}+\bar{\psi}(i\gamma^{\mu}\partial_{\mu}-m)\psi-eA_{\mu}\bar{\psi}\gamma^{\mu}\psi\Bigg)\Bigg]\,.
\label{dual.linear_z_A}
   % \end{split}
\end{align}
{ As a cross-check, it can be easily verified that integration over $\chi_{\mu\nu}$ in Eq.~(\ref{dual.linear_z_A}) reproduces the Maxwell and the $\theta$ terms with the correct coefficients.}
The Lagrangian in this partition function is linear in the gauge field $A_\mu$, which can be integrated out to produce a $\delta-$functional:
\begin{multline}
Z_{dual}=\int\big(\mathcal{D}\chi_{\mu\nu}\mathcal{D}B_{\mu\nu}\mathcal{D}\bar{\psi}\mathcal{D}\psi\mathcal{D}\Sigma_{\mu\nu}\big)\,\delta[C^{\mu}]\,\exp\Bigg[i\int d^{4}x\Bigg(-\frac{1}{4}\chi_{\mu\nu}\chi^{\mu\nu} +\frac{1}{12v^{2}}H_{\mu\nu\lambda}H^{\mu\nu\lambda}\\
-\frac{1}{2}B_{\mu\nu}\Sigma^{\mu\nu}+\bar{\psi}(i\gamma^{\mu}\partial_{\mu}-m)\psi\Bigg)\Bigg]\,,
\label{dual.Zdual1}
\end{multline}
where 
\begin{equation}
  C^{\mu}=-2\alpha\partial_{\nu}\chi^{\mu\nu}-\beta\epsilon^{\mu\nu\alpha\beta}\partial_{\nu}\chi_{\alpha\beta}-\frac{q}{2}\epsilon^{\mu\nu\alpha\beta}\partial_{\nu}B_{\alpha\beta}-e \bar{\psi}\gamma^{\mu}\psi\,.
\end{equation}
To satisfy the $\delta$  constraint, we set
\begin{equation}
2\alpha\chi_{\mu\nu}+\beta\epsilon_{\mu\nu\alpha\beta}\chi^{\alpha\beta}=-\frac{q}{2}\epsilon_{\mu\nu\alpha\beta}B^{\alpha\beta}+eJ_{\mu\nu}\,,
\label{dual.chi_equation}
\end{equation}
where we have defined the nonlocal current $J_{\mu\nu}=\frac{1}{\Box}\partial_{[\mu}j_{\nu]}$  in terms of
the fermion current  $j^\mu\equiv\bar{\psi}\gamma^{\mu}\psi\,.$ 

We contract Eq.~(\ref{dual.chi_equation}) with $\epsilon^{\mu\nu\rho\sigma}$ and solve for $\chi_{\mu\nu}$ to get
\begin{equation}
\chi_{\mu\nu}=\frac{\alpha}{2(\alpha^{2}+\beta^{2})}\Big(-q\tilde{B}_{\mu\nu}+eJ_{\mu\nu}\Big)-\frac{\beta}{2(\alpha^{2}+\beta^{2})}\Big(qB_{\mu\nu}+e\tilde{J}_{\mu\nu}\Big),
\label{dual.chisoln}
\end{equation}
where we have defined 
%
% \warning{dual means something else in this paper, let us not use it here}
%
\begin{align}
   % \begin{split}
        \tilde{B}_{\mu\nu}\equiv\frac{1}{2}\epsilon_{\mu\nu\alpha\beta}B^{\alpha\beta}\,,\\ 
        \tilde{J}_{\mu\nu}\equiv\frac{1}{2}\epsilon_{\mu\nu\alpha\beta}J^{\alpha\beta}\,.
   % \end{split}
\end{align}\label{dual.dual_defn}
%
%$\tilde{J}_{\mu\nu}$ is defined analogously. 
Substituting $\chi_{\mu\nu}$
from Eq.~(\ref{dual.chisoln}) back into the partition function (\ref{dual.Zdual1}),
we get 
\begin{align}
%    \begin{split} 
Z_{\text{dual}} = & \int\big(\mathcal{D}\chi_{\mu\nu}\mathcal{D}B_{\mu\nu} \mathcal{D}\bar{\psi}\mathcal{D}\psi\mathcal{D}\Sigma_{\mu\nu}\big) \notag\\ 
&\qquad \exp\Bigg[i\int d^{4}x\Bigg[-\frac{1}{{4}(\alpha^{2}+\beta^{2})^{2}}
\Bigg[\frac{1}{{16}}\Big(q^{2}v^{2}B_{\mu\nu}B^{\mu\nu}-e^{2}J_{\mu\nu}J^{\mu\nu}\Big) \notag\\
&\qquad\qquad +\frac{qve}{8}\tilde{B}_{\mu\nu}J^{\mu\nu}-\frac{\alpha\beta}{2}\Big(-q^{2}v^{2}B_{\mu\nu}\tilde{B}^{\mu\nu}+e^{2}J_{\mu\nu}\tilde{J}^{\mu\nu}+{qve}B_{\mu\nu}J^{\mu\nu}\Big)\Bigg] \notag \\
&\qquad\qquad +\frac{1}{12}H_{\mu\nu\lambda}H^{\mu\nu\lambda}-\frac{v}{2}B_{\mu\nu}
\Sigma^{\mu\nu}+\bar{\psi}(i\gamma^{\mu}\partial_{\mu}-m)\psi\Bigg]
\Bigg]\,,
\label{dual.dual_partition_func_3d}
 %   \end{split}
\end{align}
where we have substituted $\alpha^{2}-\beta^{2}=-\frac{1}{4}$ and rescaled $B_{\mu\nu}\rightarrow vB_{\mu\nu}$.

We choose to keep the coefficient $(\alpha^2+\beta^2)^{-2}$ intact for notational simplicity and only replace it in terms of $\theta$ in the end.
The Lagrangian presented above is the dual counterpart to the one described in Eq.~\eqref{lagrangian_original_bf}  and describes a system consisting of flux tubes whose world-sheets are described by the tensor $\Sigma^{\mu\nu}$ . These flux tubes are coupled to the two-form gauge field $B_{\mu\nu}$  which mediates the interaction between two flux tubes. Additionally, $B_{\mu\nu}$ is also coupled to a nonlocal fermion current described by the antisymmetric tensor $J^{\mu\nu}$. This kind of coupling  was first described in~\cite{choudhury2015topological} and leads to interactions between the flux tubes and fermions, mediated by the two-form gauge field. {Since the current $J_{\mu\nu}$ is nonlocal due to the $\frac{1}{\Box}$ operator, the resulting interactions will also be nonlocal. This is not unexpected --- since strings are extended objects, an interaction between fermions mediated by a string is inherently nonlocal.}

%% file: fermion_interactions.tex
\section{Effective  Interactions in the Dual Theory}\label{sec:eff.int}
%%%%%%%%%%%%%%%%%%%%%%%%%%%%%%%%%%%
Let us now calculate the effective interactions as described by the dual Lagrangian in Eq.~(\ref{dual.dual_partition_func_3d}). To get the effective interaction between flux tubes, between fermions, or between  fermions and flux tubes, we integrate out the two-form gauge field $B_{\mu\nu}$ from the path integral. Since the dual Lagrangian is quadratic in $B_{\mu\nu}$, this integral can be done exactly. In addition, since $B_{\mu\nu}$ is massive in the dual Lagrangian, a gauge fixing term not necessary. We can then segregate  the part of the partition function which depends on $B_{\mu\nu}$ and write it as
\begin{align}
%    \begin{split}
        Z_{dual}^{B} = \int\mathcal{D}B_{\mu\nu}&\exp\Bigg[i\int d^{4}x\Bigg[\frac{1}{12}H_{\mu\nu\lambda}H^{\mu\nu\lambda}+
        \frac{1}{{4}(\alpha^{2}+\beta^{2})^{2}}\Bigg(-\frac{q^{2}v^{2}}{16}B_{\mu\nu}B^{\mu\nu} -\frac{qve}{8}B_{\mu\nu}\tilde{J}^{\mu\nu}\notag\\
        &\qquad \quad 
+\frac{\theta}{32\pi^2}\Big(\frac{q^{2}v^{2}}{2}B_{\mu\nu}\tilde{B}^{\mu\nu}{-}\frac{1}{2}qveB_{\mu\nu}J^{\mu\nu}\Big)\Bigg)-\frac{v}{2}B_{\mu\nu}\Sigma^{\mu\nu}\Bigg]\,,\label{inter.partition_B.3d1}
 %  \end{split}
\end{align}
which can be recast in the form 
\begin{equation}
Z_{dual}^{B}=\int\mathcal{D}B_{\mu\nu}\exp\Bigg[i\int d^{4}x\Bigg[-\frac{1}{4}B_{\mu\nu}K^{\mu\nu\alpha\beta}B_{\alpha\beta}-\frac{1}{2}B_{\mu\nu}P^{\mu\nu}\Bigg]\Bigg]\,,
\label{inter.partition_B2.3d2}
\end{equation}
where 
\begin{equation}
K^{\mu\nu\alpha\beta}=\frac{1}{2}\Big(\Box+M^{2}\Big)g^{\mu[\alpha}g^{\beta]\nu}-{\frac{\theta}{8\pi^2}} M^{2}\epsilon^{\mu\nu\alpha\beta}B_{\alpha\beta}+\frac{1}{2}\Big(g^{\nu[\alpha}g^{\beta]\sigma}\partial_{\sigma}\partial^{\mu}-\mu\leftrightarrow\nu\Big)\,,
\label{inter.B_prop_real_space.3d}
\end{equation}          
and 
\begin{equation}
P^{\mu\nu}=\Bigg(\frac{qve}{{16}(\alpha^{2}+\beta^{2})^{2}}\tilde{J}^{\mu\nu}{+}\frac{qve\theta}{{128}\pi^2(\alpha^{2}+\beta^{2})^{2}}J^{\mu\nu}+v\Sigma^{\mu\nu}\Bigg)\,.\label{inter.current.B.3d}
\end{equation}
We have also written
\begin{equation}
    M^{2}=\frac{q^{2}v^{2}}{{16}(\alpha^{2}+\beta^{2})^{2}}=\frac{q^2v^2}{\big(1+\frac{\theta^2}{16\pi^4}\big)}\,.
\end{equation}

Integrating out $B_{\mu\nu}$ from Eq.~\eqref{inter.partition_B2.3d2} we get the following expression for the  effective interaction Lagrangian,
\begin{equation}
\mathcal{L}_{dual}^{B}=\int\frac{d^{4}k}{(2\pi)^{4}}\Bigg(-\frac{1}{4}P^{\mu\nu}(-k)G_{\mu\nu\alpha\beta}(k)P^{\alpha\beta}(k)\Bigg)\,,
\label{inter.current_inter_3d}
\end{equation}
where $P^{\alpha\beta}(k)$ is the Fourier transformation of Eq.~\eqref{inter.current.B.3d} and  $G_{\mu\nu\alpha\beta}(k)$ is the inverse of the
Fourier transformed $K^{\mu\nu\alpha\beta}$ such that 
\begin{equation}
K^{\mu\nu\alpha\beta}(k)G_{\mu\nu\alpha'\beta'}(k)=\delta_{[\alpha'}^{\alpha}\delta_{\beta']}^{\beta}\,.
\label{inter.prop.inverse.3d}
\end{equation}

The propagator is easily calculated,
\begin{align}
        G_{\mu\nu\alpha\beta}(k) &=\frac{1}{(-k^{2}+M^{2})^{2}+\frac{\theta^{2}M^{4}}{{16}\pi^{4}}}\Bigg((-k^{2}+M^{2})g_{\mu[\alpha}g_{\beta]\nu} -\frac{\theta}{{4}\pi^{2}}M^{2}\epsilon_{\mu\nu\alpha\beta}\\
        &\qquad\qquad\qquad\qquad -\Big(1-\frac{k^{2}}{M^{2}}\Big)\Big(g_{\nu[\alpha}g_{\beta]\sigma}k^{\sigma}k_{\mu}-\mu\leftrightarrow\nu\Big)\Bigg)\,,
    \label{inter.prop.B.3d.kspace.theta}
\end{align}
where we have used Eq.~\eqref{dual.alpha_beta_soln} to substitute for the value of $\alpha\beta\,.$

From Eqs.~\eqref{inter.current.B.3d}, \eqref{inter.current_inter_3d}, and \eqref{inter.prop.B.3d.kspace.theta}, we
can calculate the effective interactions, which can be split into three parts as follows:
\begin{align}
%    \begin{split}
      \mathcal{L}^B_{\text{dual}}=& \int\frac{d^{4}k}{(2\pi)^{4}}\left(\frac{qve}{16(\alpha^{2}+\beta^{2})^{2}}\tilde{J}^{\mu\nu}(-k)+\frac{qve\theta}{128\pi^2(\alpha^{2}+\beta^{2})^{2}}J^{\mu\nu}(-k)\right) G_{\mu\nu\alpha\beta}(k)\notag \\
        &\qquad\qquad\qquad  \times\left(\frac{qve}{16(\alpha^{2}+\beta^{2})^{2}}\tilde{J}^{\alpha\beta}(k)+
        \frac{qve\theta}{128\pi^2(\alpha^{2}+\beta^{2})^{2}}J^{\alpha\beta}(k)\right) \notag \\
        &\qquad +v^2\int\frac{d^{4}k}{(2\pi)^{4}}\Sigma^{\mu\nu}(-k) G_{\mu\nu\alpha\beta}(k)
        \Sigma^{\alpha\beta}(k)
        +\cdots
%    \end{split}
    \label{inter.inter.all.1.3d}
\end{align}
The first integral in the above equation corresponds to the effective fermion-fermion interaction. The second integral describes the interaction between flux tubes while the ellipses represent the $J-\Sigma$ interaction terms.

Evaluating the first integral, we get
\begin{align}
%    \begin{split}
\frac{e^{2}}{4(\alpha^{2}+\beta^{2})^{2}}\int\frac{d^{4}k}{(2\pi)^{4}}\Bigg[j_{\mu}(-k)\Bigg[\frac{M^{2}}{(-k^{2}+M^{2})^{2}+\frac{\theta^{2}M^{4}}{4\pi^{4}}}\frac{1}{k^{2}}\Big(k^2-M^2 + \frac{\theta^2}{128\pi^4}\big(k^2+7M^2\big)\Big)\Bigg]j^{\mu}(k)\Bigg]\,.
 %          \end{split}
           \label{inter.fermion.fermion.3d}
\end{align}
The above expression gives the effective interaction between two fermions
in the system. Clearly, in the presence of a non-zero $\theta$ term,
the interaction is different from that of a Coulombic or a Yukawa
interaction. In addition, it is clear that $\theta$ also contributes
to the mass of the $B$ field. It is also interesting to note that
$\theta$ and $M$ appear together in this expression. 
%Letting $M\rightarrow 0$\,, we find  that then $\theta$  has no effect on the interactions suggesting
%that the axion term becomes important for fermionic interactions in the presence of a background of a charged massive scalar condensate. 
%
%\warning{{$M\rightarrow 0$ is not a good limit in presence of strings. We have to say things differently.}}
Thus the effect of the axion term on fermionic interactions depends on the scale at which condensation occurs, as well as the charge of the condensate.
It is important to note here that the axion term contributes to the dynamics of fermions in the bulk despite being a topological term and it also shifts the pole of the propagator. The position of the pole in the complex plane now depends on the value of $\theta$\,. Additionally, as $\theta\rightarrow 0$, the resulting effective interaction between fermions reduces to
\begin{align}
%\begin{split}
     & \int\frac{d^{4}k}{(2\pi)^{4}}\frac{e^{2}}{2(\alpha^{4}+\beta^{4})}\Bigg[j_{\mu}(-k)\Bigg(\frac{1}{k^{2}-M^{2}}-\frac{1}{k^{2}}\Bigg)j^{\mu}(k)\Bigg]\,,
 % &\qquad \qquad 
 % =32\int\frac{d^{4}k}{(2\pi)^{4}}e^{2}\Bigg[j_{\mu}(-k)\Bigg(\frac{1}{k^{2}-M^{2}}-\frac{1}{k^{2}}\Bigg)j^{\mu}(k)\Bigg]\,.\label{inter.fermion.fermion.zero.theta.3d}
%\end{split}
\end{align}
which agrees  with the result obtained in~\cite{MUKHERJEE2020168167} where the
$\theta$ term was not considered. 

We will now focus on the second integral in Eq.~\eqref{inter.inter.all.1.3d}, which is
\begin{align}
%    \begin{split}
        2v^{2}\int\frac{d^{4}k}{(2\pi)^{4}}\Sigma_{\mu\nu}(-k)\left[\frac{1}{(-k^{2}+M^{2})^{2}+\frac{\theta^{2}M^{4}}{16\pi^{4}}}\left((-k^{2}+M^{2})\Sigma^{\mu\nu}(k)-\frac{\theta M^{2}}{2\pi^{2}}\tilde{\Sigma}^{\mu\nu}(k)\right)\right]\,.
%    \end{split}
    \label{inter.vortex.vortex.3d}
\end{align}
As stated before, this term gives the effective interaction between two flux tubes in the system. It can be seen that the effect of the axion term is two-fold. Firstly, it has shifted the pole of the propagator of the gauge field $B_{\mu\nu}$. And secondly, there is a new kind of interaction between the world-sheet tensor $\Sigma_{\mu\nu}$ and its dual which is generated purely because of the axion term.

To obtain the fermion-flux tubes interactions, we evaluate the terms represented by the ellipses in Eq.~(\ref{inter.inter.all.1.3d})\,. Remembering the  gauge invariance condition $\partial_\mu\Sigma^{\mu\nu} =0$\,, we get 
\begin{align}
 %   \begin{split}
        \frac{4eM^{2}}{q}\int\frac{d^{4}k}{(2\pi)^{4}}\;j_{\nu}(k)\,\frac{1}{(-k^{2}+M^{2})^{2}+\frac{\theta^{2}M^{4}}{16\pi^{4}}}\Bigg[-\frac{k_{\mu}}{k^{2}}\Big(-k^{2}+M^{2}\Big(1+\frac{\theta^{2}M^{2}}{8\pi^{2}}\Big)\Big)\Bigg]\tilde{\Sigma}^{\mu\nu}(-k)\,.
    \label{inter.fermion.vortex.3d}
\end{align}
It can be seen again that the effect of the axion term is  {to change} the interaction between the fermion current $j_\mu$  and the dual of flux tube given by $\Sigma^{\mu\nu}$. The interaction between these two fields in the absence of the $\theta$  term was obtained in \cite{MUKHERJEE2020168167}. 
 We note that all of these interactions, fermion-fermion, flux tube-flux tube and the fermion-flux tube have the common feature of having a term proportional to $\theta M^2$ which means that the axion term has an observable effect when there is a charged massive scalar condensate in the background. {While the physical picture behind fermions interacting via a string is clear, we will illustrate an important physical consequence of the interaction between the fermions and flux strings in $2+1$ dimensions. This is the phenomenon of flux pinning to a fermion and resulting anyonic statistics.}

%% file: dimensional_reduction.tex
\section{Dimensional Reduction and Effective (2+1)D Theory} 
\label{sec:dim_red}
%%%%%%%%%%%%%%%%%%%%%%%%%%%%%%%%%%%%%%%%%%%%%%%%%%%%%

In the previous sections, we have shown how the presence of a nonzero $\theta$ term affects local interactions in the presence of a massive scalar condensate. It was shown in~\cite{mukherjee2022spin}, albeit for vanishing $\theta$, that magnetic flux attachment to electrons present in the system is obtained when the theory is dimensionally reduced to 2+1  dimensions. {To understand the pinning of magnetic flux on a particle, one can recall the Chern-Simons(CS) theory in $2+1$ dimensions coupled to a charge current~\cite{ZHKModel,fieldtheoryofanyons,ZHANG1992}. For convenience, if we consider a point charge distribution with the charge density given by $\rho \sim \sum_{i}\delta(\vec{x}-\vec{x}_i)$, we find that the magnetic field in Chern-Simons theory is present precisely at the location of charges. One can therefore look at this as if the charges were carrying a point magnetic flux with them. Because of this flux, the Aharonov-Bohm (AB) phase obtained by the wavefunction of two such composite particles when they are exchanged is dependent on the relative flux carried by them. The resulting AB phase upon exchange is neither 0 i.e. bosonic nor $\pi$ i.e. fermionic. Hence, such particles exhibit anyonic statistics.
 Motivated by the results of \cite{mukherjee2023spin}, we show here that such a flux pinning can be obtained as a consequence of the fermion-flux tube interaction presented in the previous section concluding with Eq.~(\ref{inter.fermion.vortex.3d}). The flux pinning here will however be due to the charged particles getting attached to the point-like flux passing through the vortices. As shown in Eq.~(\ref{inter.fermion.vortex.3d}), the fermion-vortex interaction is $\theta$ dependent, we will find here as well that the flux that gets pinned to the fermions will also be $\theta$ dependent.}

We follow the steps outlined in~\cite{mukherjee2023spin} to perform a dimensional reduction of our system. {We consider our system to have finite extension $L$ along the $Z$ direction. We take $L$ to be much smaller than the size of the system in the other two directions, consequently making the system quasi $2d$.}
In addition, we also assume that all the fields in the Lagrangian are independent of the third coordinate $Z$.  The last step is necessary because it allows us to separate the Lagrangian into a $2+1$ dimensional part and a $1d$ part.
We begin by segregating the $(0,1,2)$ components from the 3 component of all the fields in the Lagrangian. 
\begin{align}
-\frac{1}{4}F^{\mu\nu}F_{\mu\nu} &=-\frac{1}{4}F_{ab}F^{ab}+\partial_{a}\varphi\partial^{a}\varphi, \\
\bar{\psi}(i\gamma^{\mu}\partial_{\mu}-m)\psi-eA_{\mu}j^{\mu} &=\bar{\psi}(i\gamma^{a}\partial_{a}-m)\psi-eA_{a}j^{a}-e\varphi j^{3}\,,  \\
% %
\frac{1}{12}H^{\mu\nu\lambda}H_{\mu\nu\lambda} &=-\frac{1}{4}H^{ab}H_{ab}+\frac{1}{12}H^{abc}H_{abc}\,,  \\
% %
B^{\mu\nu}B_{\mu\nu} &=2B_aB^a + B_{ab}B^{ab}\,,
\label{dim.red.expansion}
\end{align}
where we have written $\epsilon_{abc}=\epsilon_{abc3}$ and $\varphi=A_z$. Additionally, the wave operator $\Box=\partial_\mu\partial^\mu$  now takes the form $\Delta=\partial_a\partial^a$ .
The Latin indices $a,b,c, \cdots$ represent the $(2+1)$-dimensional components and take the values $0,1,2$. {Reduction of the fields to a $2+1$ dimensional spacetime changes the mass dimensions of the fields and the coupling constants. Because of this the fields and the coupling factors pick up a scaling factor as a power of $L$. We have therefore rescaled the fields appropriately in the action inside the partition function below and absorbed them in the redefinition of the fields, e.g., $A_{a}\rightarrow\sqrt{L}A_{a}$.} Substituting the above expressions into the dual partition function, we get \\
\begin{align}
   Z_{\text{dual}}=\int & 
 { \mathcal{D}B_{ab}\mathcal{D}B_{a}\mathcal{D}\bar{\psi}\mathcal{D}\psi\mathcal{D}\Sigma_{ab}\mathcal{D}\Sigma_{a}\mathcal{D}A_{a} \mathcal{D}\varphi}
 \notag\\
 %  \big(\mathcal{D}\chi_{\mu\nu}\mathcal{D}B_{\mu\nu}\mathcal{D}\bar{\psi}\mathcal{D}\psi\mathcal{D}\Sigma_{\mu\nu}\mathcal{D}A_{\mu}\big)
  % \notag\\
&\exp\Bigg[i\int d^{4}x\Bigg[-\frac{1}{4}F_{ab}F^{ab}+\partial_{a}\varphi\partial^{a}\varphi+\bar{\psi}(i\gamma^{a}\partial_{a}-m)\psi-\kappa\tilde{e}A_{a}j^{a}-\kappa\tilde{e}\varphi j^{3}\Bigg] \notag \\
&\qquad+\frac{1}{{4}(\alpha^{2}+\beta^{2})^{2}}\Bigg[-\frac{q^{2}v^{2}}{16}\left(2B_{a}B^{a}+B_{ab}B^{ab}\right){-}\frac{qve}{4}\left(\tilde{B}_{a}\frac{1}{\Delta}\partial^{a}j^{3}+\frac{1}{2}\epsilon_{abc}B^{a}\partial^{b}\frac{1}{\Delta}j^{c}\right)\notag \\
&\qquad +\frac{\theta}{32\pi^2}\Big({+}4q^{2}v^{2}\tilde{B}_{c}B^{c}+4qve\left(\partial^{a}B_{a}\frac{1}{\Delta}j^{3}{-}\partial^{a}B_{ab}\frac{1}{\Delta}j^{b}\right)\Big)\Bigg] \notag \\
&\qquad -\frac{1}{4}H^{ab}H_{ab}+\frac{1}{12}H^{abc}H_{abc}-vB_{a}\Sigma^{a}-\frac{v}{2}B_{ab}\Sigma^{ab}\Bigg]\,,
 \label{dim.red.dual.1}
\end{align}
{where we have used the following relation to restore the photon field $A_\mu$ in Eq.~(\ref{dual.dual_partition_func_3d}):
\begin{equation}
    \int \mathcal{D}A_\mu \exp\left(i\int d^4x\left(-\frac{1}{4}F_{\mu\nu}F^{\mu\nu}+A_\mu K^\mu\right)\right)\propto\int\exp\left(-\frac{i}{2}\int d^4x \left(K_\mu\frac{1}{\Box}K^\mu\right)\right)\,.
\end{equation}}

We consider the special case when the fermion current is confined to the $X-Y$ plane, and set $j_3=0$, to get 
\begin{align}
  %  \begin{split}
Z_{\text{dual}}\Big|_{j_3=0}=&\int
\mathcal{D}B_{ab}\mathcal{D}B_{a}\mathcal{D}\bar{\psi}\mathcal{D}\psi\mathcal{D}\Sigma_{ab}\mathcal{D}\Sigma_{a}\mathcal{D}A_{a} {\mathcal{D}\varphi} \notag\\
&\exp\Bigg[i\int d^{4}x\Bigg[-\frac{1}{4}F_{ab}F^{ab}+\partial_{a}\varphi\partial^{a}\varphi+\bar{\psi}(i\gamma^{a}\partial_{a}-m)\psi-\kappa\tilde{e}A_{a}j^{a} \notag\\
&\qquad\qquad  -\frac{{M^2}}{{4}}\left(2B_{a}B^{a}+B_{ab}B^{ab}\right){-}\frac{qv{\kappa\tilde{e}}}{{2}}\epsilon_{abc}B^{a}\partial^{b}\frac{1}{\Delta}j^{c} \notag\\
&\qquad\qquad +{\frac{\theta M^2}{2\pi^2}}\tilde{B}_{c}B^{c}{-}{\frac{{2}\theta \kappa\tilde{e}}{\pi^2}}\partial^{a}B_{ab}\frac{1}{\Delta}j^{b}\notag \\
&\qquad\qquad -\frac{1}{4}H^{ab}H_{ab}+\frac{1}{12}H^{abc}H_{abc}-vB_{a}\Sigma^{a}-\frac{v}{2}B_{ab}\Sigma^{ab}\Bigg]\,,
\label{dim.red..z.j3.0}
  %  \end{split}
\end{align}
where $\kappa = L^{-\frac{1}{2}}$ and the ``screened'' charge $\tilde{e}$ is defined as
\begin{equation}
  \tilde{e} = \frac{e}{{16}(\alpha^{2}+\beta^{2})}\label{e_tilde}=\frac{Me}{2qv}\,.
\end{equation}
We wish to integrate out the $B_a$ and the $B_{ab}$ fields from the partition function to obtain the effective electrodynamics on the thin layer. To this end, we integrate the two fields separately, treating the mixed term ${\frac{\theta M^2}{2\pi^2}\tilde{B}_cB^c}$ perturbatively. We begin by writing out the terms in the partition function containing the relevant fields \\
%\centerline{\color{purple}\bf Check parentheses!}
%
\begin{align}
 %   \begin{split}
       Z_{\text{dual}}^B\Big|_{j_3=0}=&\int\big(\mathcal{D}B_{ab}\mathcal{D}B_{a}\big) \notag \\
& \exp\Bigg[i\int d^{3}x\Big(-\frac{1}{4}H^{ab}H_{ab}-\frac{M^2}{2}B_a B^a{-\frac{qv\kappa\tilde{e}}{2}}\epsilon_{abc}B^{a}\partial^{b}\frac{1}{\Delta}j^{c}-vB_{a}\Sigma^{a} \notag \\
&\qquad\qquad +\frac{1}{12}H^{abc}H_{abc}-\frac{M^2}{2}B_{ab}B^{ab}-{\frac{2\theta \kappa\tilde{e}}{\pi^2}}B_{ab}\partial^{a}\frac{1}{\Delta}j^{b}+B_{ab}\Sigma^{ab} \notag \\
&\qquad\qquad +\frac{\theta M^2}{2\pi^2}\tilde{B}_{c}B^{c}\Big)\Bigg]\,.
 \label{dim.red..z.j3.0_combined}
 %   \end{split}
\end{align}

The propagators for $B_a$ and $B_{ab}$ are respectively
\begin{align}
   % \begin{split}
        \Delta^0_{ab}=\frac{1}{-k^{2}+M^{2}}\Bigg(g_{ab}-\frac{k_{a}k_{b}}{M^{2}}\Bigg)\,,
   % \end{split}
    \label{eq: Ba_prop}
\end{align}
and 
\begin{align}
   % \begin{split}
        \Delta^0_{ab,cd}=\frac{1}{-k^2+M^2}\Big[\big(g_{ac}g_{bd}-g_{ad}g_{bc}\big)-\frac{g_{ac}k_bk_d-g_{ad}k_bk_c}{M^2}\Big]\,.
   % \end{split}
    \label{eq: Bab_prop}
\end{align}
The total propagator for $B_a$ including the mixed term can be calculated from an infinite sum following~\cite{Allen:1990gb},
\begin{align}
  %  \begin{split}
        \Delta_{ab}=\Delta^0_{ab}+\Delta^0_{ac}\Theta_{cd}\Delta^0_{db} + \Delta^0_{ac}\Theta_{cd}\Delta^0_{de}\Theta_{ef}\Delta^0_{fb}+\cdots
  %      \text{. . . . .}
  %  \end{split}
\end{align}
where 
\begin{align}
%    \begin{split}
        \Theta_{ab}=V_{a,cd}\Delta^0_{cd,ef}V_{ef,b}\,,
%    \end{split}
\end{align}
with $V_{ab,c}=V_{c,ab}=\frac{\theta M^2}{2\pi^2}\epsilon_{abc}$\,. { Note that the source terms have to be ignored while calculating the propagator. }
The total propagator is then 
\begin{align}
 %   \begin{split}
        \Delta_{ab}=-\frac{g_{ab}-\frac{k_ak_b}{M^2}}{k^2-(M^2+\mu^2)}\,,
 %   \end{split}
\end{align}
where 
\begin{align}
 %   \begin{split}
        {\mu^2=\frac{\theta^2 M^2}{8\pi^4}}\,.
%    \end{split}
\end{align}
Having found the effective propagator for $B_a$, we can write down the vortex-fermion and the fermion-fermion interaction 
by integrating out the $B$ fields from the partition function of Eq.~(\ref{dim.red..z.j3.0}), 
%from the following partition function 
%
\begin{align}
  Z_{\text{dual}}^B\Big|_{j_3=0}
\propto \int
\exp\left[-\frac{i}{2}\int d^{3}x d^3y\Big(P_a(x)\Delta^{ab}(x-y)P_b(y)\Big)\right]\,,
\label{dim.red.P-Delta-P}
 \end{align}
where we have written 
\begin{align}
    P_a(x) =-\frac{qv\kappa\tilde{e}}{2}\epsilon_{abc}\partial^{b}\frac{1}{\Delta}j^{c}-v\Sigma_{a}\equiv
-\frac{qv\kappa\tilde{e}}{2}\tilde{j}_a-v\Sigma_{a}\,.
\end{align}

The vorticity $\Sigma^a$ can be split into two parts, one coming from the singular part of the scalar field $\phi$ that we started with in 
Eq.~(\ref{linear_z_A}) and another coming from any external magnetic field that is applied, hence 
$\Sigma^a = \epsilon^{abc}\partial_b \partial_c\chi^s + \Sigma_{\rm e}^a\equiv \epsilon^{abc}\partial_b S_c + \Sigma_{\rm e}^a$\,.
The partition function in Eq.~(\ref{dim.red.P-Delta-P}) can then be rewritten as 
\begin{align}
%    \begin{split}
\int
\exp \left[-\frac{i}{2}\int \right. &d^{3}x d^3 y  \left(\frac{q^2v^2\kappa\tilde{e}^2}{4}\tilde{j}^a\frac{1
}{\Delta+(M^2+\mu^2)}\tilde{j}_a + qv^2\kappa\tilde{e}S^d \frac{1
}{\Delta+(M^2+\mu^2)} \epsilon_{bcd} \partial^c \tilde{j}^b \right. \notag\\
&+\frac{v^2}{2}G_{ab}\frac{1}{\Delta^2 + (M^2 + \mu^2)}G^{ab} + v^2 \Sigma^a_{\rm e}
\frac{1}{\Delta^2 + (M^2 + \mu^2)}\Sigma_{{\rm e}a} \notag \\
& \left.\left.+ v^2 \epsilon^{abc}G_{bc}\frac{1}{\Delta^2 + (M^2 + \mu^2)}\Sigma_{{\rm e}a} + qv^2\kappa\tilde{e}
\Sigma^a_{\rm e} \frac{1}{\Delta^2 + (M^2 + \mu^2)} \tilde{j}_a\right)\right]\,,
%    \end{split}
\end{align}
where we have written $G_{ab} = \partial_{a} S_{b} - \partial_b S_a\,.$ Collecting the terms with the field $S^a$, and using the fact that at low energies $\dfrac{1}{\Delta^2 + (M^2 + \mu^2)}\approx\dfrac{1}{M^2+\mu^2}$\,, we get for the vortex interaction 
\begin{align}
\int
\exp\Bigg[-\frac{i}{2}\int d^{3}x & \Big[ qv^2\kappa\tilde{e} S^d \frac{1
}{(M^2+\mu^2)} \epsilon_{bcd} \partial^c \tilde{j}^b +\frac{v^2}{2}G_{ab}\frac{1}{(M^2 + \mu^2)}G^{ab} \notag \\
&\qquad + v^2 \epsilon^{abc}G_{bc}\frac{1}{ (M^2 + \mu^2)}\Sigma_{{\rm e}a} \Big]\,.
\end{align}
We rescale $S^a\rightarrow\frac{v}{\sqrt{M^2+\mu^2}}{S^a}$ and extract the Lagrangian from the above partition function as
\begin{equation}
    \mathcal{L}_S = -\frac{1}{4}G_{ab}G^{ab} + S_c\Bigg(\frac{qv\kappa\tilde{e}}{\sqrt{M^2+\mu^2}}j^c+\frac{v}{\sqrt{M^2+\mu^2}}\epsilon^{abc}\partial_a\Sigma_{{\rm e}b}
    \Bigg)\,.\label{flux_pinning}
\end{equation}
Following~\cite{mukherjee2023spin}, we can say that the equation of motion derived from the above Lagrangian shows that 
the number of vortices present in the system, including those due to the external magnetic field, is exactly the same as the charge,
\begin{equation}
    \epsilon^{bc}\partial_b S_c  + \frac{v}{q\sqrt{M^2+\mu^2}}\Sigma_{{\rm e}0}= -\frac{v\kappa\tilde{e}}{\sqrt{M^2+\mu^2}}\tilde{j}_0\,.
\end{equation}
{It is instructive to understand this equation by comparison with that of the Chern-Simons theory. We digress briefly here to write down the equations of motion of CS theory coupled to a charge current. The Lagrangian is given by 
\begin{equation}
    L_{CS} = \frac{k}{4\pi}\epsilon^{\mu\nu\alpha}\partial_\mu A_\nu A_\alpha +A_\mu j^\mu\,,
    \label{eq:CS Lagrangian}
\end{equation}
and the resulting equations of motion are
\begin{align}
     \epsilon^{ab}\partial_aA_b &= -\frac{2\pi}{k}j^0 \,, \\
     \epsilon^{ab}\partial_0A_b &= -\frac{2\pi}{k}j^a\,.
\end{align}

We will need only the first of these equations of motion for our purposes. On the left hand side of the equation we have the magnetic field $B$ which in $2d$ points in the $\hat{z}$ direction. This magnetic field, which is a function of position $(x,y)$, is proportional to the charge density $\rho=j^0$ at that point.  Therefore we say that the magnetic flux is pinned on the charge. The number $k$, also called the \textit{level} of the CS theory appears as the coefficient of proportionality and determines the exchange statistics of the charged particles.

Coming back to Eq.~(\ref{flux_pinning}), the first term on the left hand side is the vorticity due to the singular flux points formed because of the symmetry breaking and formation of the scalar condensate, and the second term is vorticity arising in the system if an external magnetic field is present. These are both functions of the position $(x,y)$. On the right hand side we have a term proportional to the charge density with a $\theta$-dependent coefficient. Hence, this equation describes the attachment of vorticity, and consequently the flux of the vortices to the charges present in the system. As in the case of the CS theory, the exchange statistics of these charged particles will be anyonic, however, here the statistics are determined by the parameters $M$ and $\theta$.}

%% file: summary.tex
\section{Summary and outlook}
%%%%%%%%%%%%%%%%%%%%%%%%%%%%%%%%%%%%%%%%%%
In this paper, we have considered vortex strings in an Abelian Higgs model coupled to $\theta$-electrodynamics. In addition, we have taken the electromagnetic field to be also coupled to charged fermions. Treating this system as a four-dimensional relativistic field theory, we have dualized the fields to write the path integral in terms of a rank-2 tensor field $B_{\mu\nu}$ which couples to the itinerant fermions via a nonlocal interaction term. The emergence of the 2-form gauge field $B_{\mu\nu}$ is not unexpected, as it is well known that such a field mediates the interaction between vortex strings. The nonlocal interaction, although less well known, emerges in a straightforward manner from the path integral of the model. Using these, we calculate the effective interaction between fermions, between strings, and between fermions and strings. These may be relevant to type-II superconductors with itinerant fermions. 

We find that the parameter $\theta$ appears in all of these interactions, modifying the pole of the propagator, as well as the overall strength of the interaction. Since the axion term is a total divergence, it is not expected to affect  local physics. It is thus a surprise that, as we have shown in this paper, the presence of vortex strings and itinerant fermions in the system brings the effect of the axion term down to local dynamics. This is our main result. Intuitively, we can understand this by realizing that the existence of the string creates a circle at infinity from which the total divergence term gets a nontrivial contribution for the path integral. However, we are not aware of any method of seeing this contribution in local dynamics other than the dualization procedure used in this paper. This result also justifies our use of 3+1-dimensional relativistic quantum field theory to analyze this system, even though we may be thinking of its application in superconductors, as the dualization is specific to 3+1-dimensional relativistic field theories. {It is interesting to look at the $\theta$ term from the perspective of the ABJ anomaly~\cite{FujikawaABJanomaly, Adler, BellJackiw}. In the usual case, the anomaly appears for massless fermions, while for massive fermions one can use the anomaly calculation to trade the $\theta$ term for a phase of the fermion mass $m$. However, such a calculation is problematic in the presence of singular gauge configurations. We can see this by noting that in Fujikawa's method, the fermions are expanded in the basis of the eigenvectors of the Dirac operator $i\slashed{D}$\,, which becomes ill-defined at the location of the singularities.} 

%it is important to note that the singularities present in the gauge field configuration for the case we consider makes such a calculation problematic.}

We have also considered a dimensionally reduced version of our system, corresponding to all the fields being restricted to a thin layer. We know that in the absence of the axion term, the reduced system shows that charges are attached to the vortex strings in a kind of flux attachment. Our calculation with the axion term finds the same result --- the density of vortices is exactly the same the density of the (dual) charge. It should be possible, using the same techniques, to calculate the effective interactions between vortices in thin slabs of type-II superconductors when an axion term is present. We leave these and similar questions for future work.

%% file: acknowledgements.tex
\section*{Acknowledgements}
S.K. thanks Department of Science and Technology, India for a Junior Research Fellowship, and  Shantonu Mukherjee for helpful discussions.